\documentclass{vgtc}                          

\ifpdf
  \pdfoutput=1\relax                   
  \usepackage{graphicx}                
  \DeclareGraphicsExtensions{.pdf,.png,.jpg,.jpeg} 
\else
  \usepackage{graphicx}                
  \DeclareGraphicsExtensions{.eps}     
\fi%

\graphicspath{{figures/}{pictures/}{images/}{./}} 
\usepackage[shortcuts]{extdash}
\usepackage{microtype}                 
\PassOptionsToPackage{warn}{textcomp}  
\usepackage{textcomp}                  
\usepackage{mathptmx}                  
\usepackage{times}                     
\usepackage{tabularx}
\usepackage{cite}                      
\usepackage{tabu}                      
\usepackage{booktabs}                  
\usepackage{caption}
\usepackage{subcaption}
\usepackage{enumitem}
\usepackage{setspace}

\onlineid{0}

\vgtccategory{Research}

\title{Extending the Metaverse: Hyper-Connected Smart Environments with Mixed Reality and the Internet of Things}

\author{Jie Guan \thanks{jie.guan@ocadu.ca} %
\and Alexis Morris  \thanks{amorris@ocadu.ca} %
\and Jay Irizawa\thanks{jirizawa@ocadu.ca}}
\affiliation{\scriptsize  Adaptive Context Environment (ACE) Lab \\OCAD University}

\teaser{
\centering
 \includegraphics[width=\linewidth]{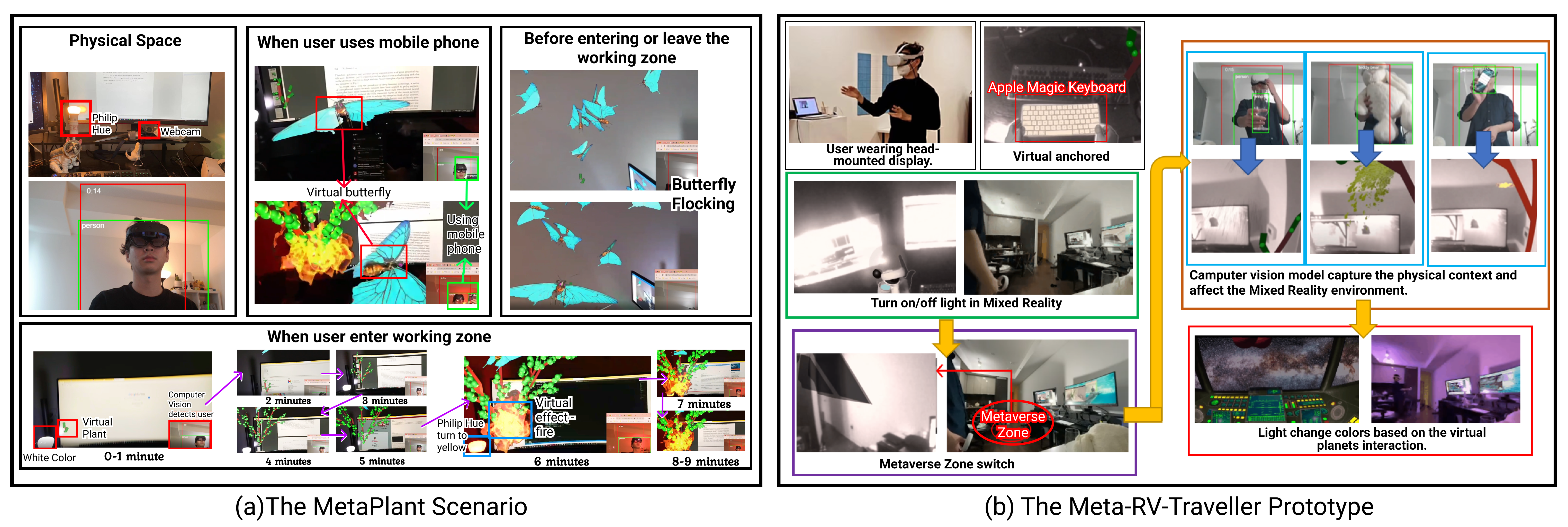}
 \caption{Extended Metaverse prototypes, as in \cite{guan2022extendingThesis}: (a)The prototype showing connection between physical and virtual environment. A growing tree represents time-context, and a merger of visual effects, and physical lighting dynamics enrich the environment. (b)Prototype for Transitioning in a Hyper-Connected Metaverse Environment. The Meta-RV-Traveller prototype explores possible user interactions in the hyper-connected metaverse, as the user transitions between virtual, mixed, and physical environments.}
 \label{Prototypes}
}

\abstract{The metaverse, i.e., the collection of technologies that provide a virtual twin of the real world via mixed reality, internet of things, and others, is gaining prominence. However, the metaverse faces challenges as it grows toward mainstream adoption. Among these is the lack of strong connections between metaverse objects and traditional physical objects and environments, which leads to inconsistencies for users within metaverse environments. To address this issue, this work explores the design and development of a framework for bridging the physical environment and the metaverse through the use of internet-of-things objects and mixed reality designs. The contributions of this include: i) an architectural framework for extending the metaverse, ii) design prototypes using the framework. Together, this exploration charts the course toward a more cohesive and hyper-connected metaverse smart environment.

} 

\CCScatlist{
 Virtual Reality, Mixed Reality, Augmented Reality, Extended Reality, Internet of Things, Human Computer Interaction}

\begin{document}
\begin{titlepage}

     \vspace{1cm}
        Full Citation: J. Guan, A. Morris and J. Irizawa, "Extending the Metaverse: Hyper-Connected Smart Environments with Mixed Reality and the Internet of Things," 2023 IEEE Conference on Virtual Reality and 3D User Interfaces Abstracts and Workshops (VRW), ShangHai, China, 2023, pp. 817-818, DOI: 10.1109/VRW58643.2023.00251.

       \vspace*{1cm}

       \copyright2023 IEEE. Personal use of this material is permitted.  Permission from IEEE must be obtained for all other uses, in any current or future media, including reprinting/republishing this material for advertising or promotional purposes, creating new collective works, for resale or redistribution to servers or lists, or reuse of any copyrighted component of this work in other works.

       \vspace{1.5cm}
  
\end{titlepage}

\firstsection{Toward an Extended Metaverse: Solving the Metaverse Disconnect with XR-IoT Designs}

\maketitle
\begin{figure}[tbh]
 \centering 
 \includegraphics[width=\linewidth]{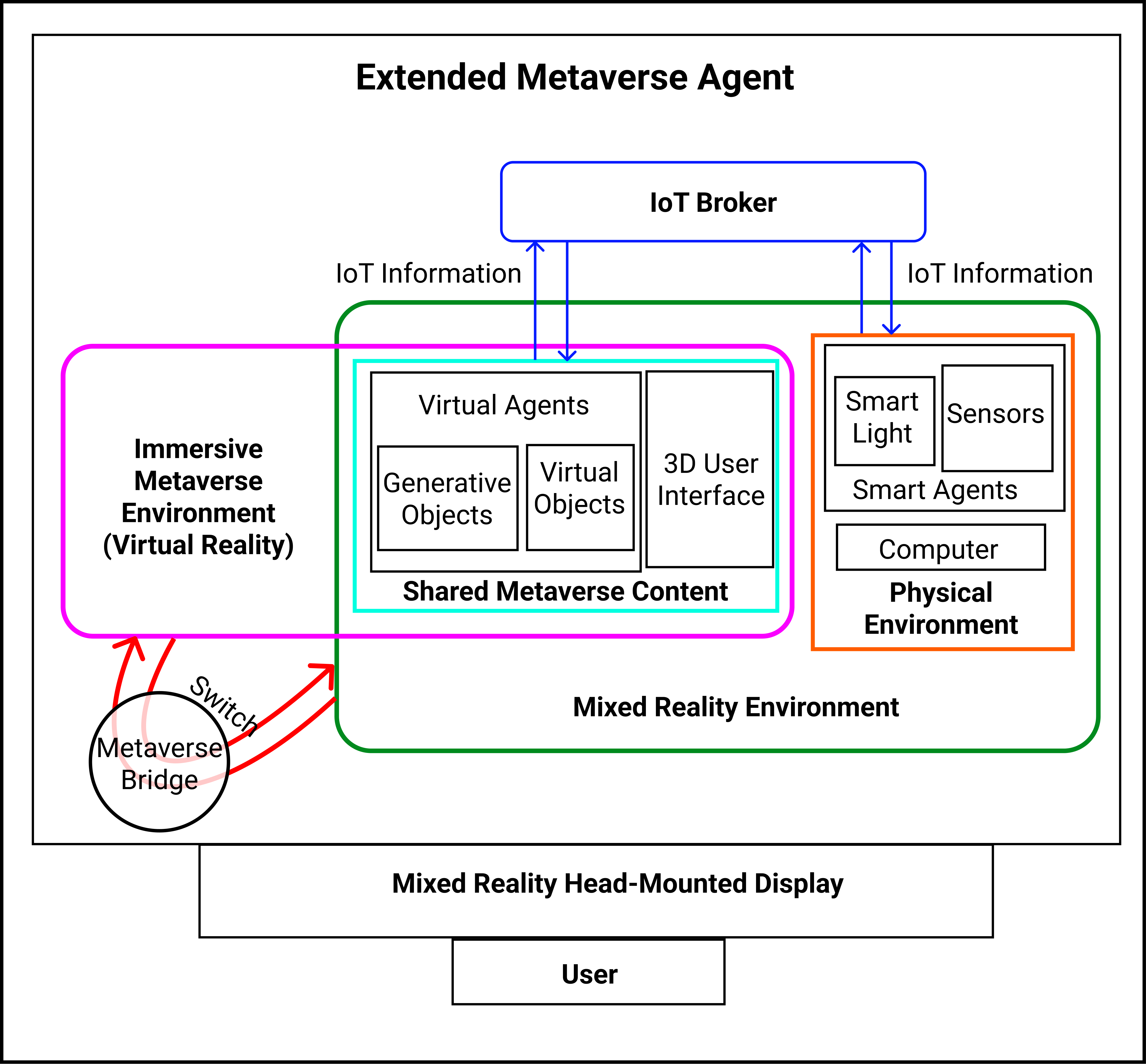}
 \caption{Conceptual Architecture for extended metaverse framework systems that strongly connect the real and the virtual (i.e., the entire RV spectrum \cite{Milgram2011}). A metaverse bridge enables switching between immersive VR and XR environments while an IoT broker enables communication and control of IoT physical and virtual objects.}


 \label{Architecture}
\end{figure}

Humans are inherently anchored to physical spaces, and must interact with the physical environment continuously, for instance to eat, rest, manage their surroundings, and other social interactions within their physical environments. As a result, when the human within a metaverse environment is fully immersed and the metaverse environment is not connected to the dynamics of the physical environment this introduces significant information gaps, or a  disruption of signals as "noise" between physical and virtual space as described by Shannon and Weaver's model of communication \cite{Shannon1964}. As the level of immersiveness within the virtual experience increases, this disconnect can also be expected to increase. Consequently, if the metaverse does not have the means to maintain the coherence with the physical world, then the many new metaverse applications will be limited to virtual human-environment activities while increasingly disconnected from physical human-environment activities. Hence, the overloading of processing information in the metaverse without rules and limitations from the physical world, or a proper design of the connection between them, may limit metaverse simulated environments. For example, when users are wearing full virtual reality head-mounted displays, they tend to be largely unaware of their physical surroundings, leading to unexpected interactions (such as bumping into objects, missed social awareness cues, missed time cues, and other context awareness factors).

Merging the literature on the metaverse, XR-IoT(XRI), context awareness, Mixed Reality Agents, presence and generative design is considered in this work as a path toward addressing the metaverse disconnect problem\cite{Guan2022IEEEVR}\cite{Morris2021}\cite{Tsang2021}. This is synthesized into the following architecture and scene design in Figure \ref{Architecture}.

\textbf{Architecture Component Design:} In the architecture (see Figure \ref{Architecture}), it provides users with an immersive extended metaverse experiment visually, an HMD headset is considered as the primary mixed reality platform, as opposed to handheld MR. The design of extended metaverse agent contains three main components, including mixed-reality environment (with two sub-components, shared metaverse content and physical environment), the immersive metaverse environment (also including shaed metaverse content) which represents the virtual reality, and the IoT broker as the bridge. The system enables users to transition across these hybrid spaces, as shown by the red arrows, indicating users could switch between immersive metaverse environment (virtual reality) and a mixed reality environment. In addition, the blue arrows highlight the bi-directional information communication through IoT broker between shared metaverse content and the physical environment.


\textbf{Bi-directional Connections and Transitions in Hyper-Connected Metaverse Environment:} The architecture(see Figure \ref{Architecture}) showed a strong connectedness between shared metaverse content and physical environment through IoT broker (presented as blue box) and IoT information (showed as blue arrows). In addition with the transitions and switch (showed as red arrows) between immersive metaverse environment and the mixed reality environment through metaverse bridge, a user is able to seamlessly experience the dynamics of the real while in the fully virtual and the dynamics of the fully virtual while in either mixed reality or (transitively) the fully physical environment. These features of the architecture targets the disconnect challenge directly, allowing for users to easily become aware of both the physical and the virtual surrounding context and dynamics, via IoT connectivity (blue arrows and box), and visual transitions in mixed reality and immersive metaverse through metaverse bridge (red arrows).

\section{Frameworks and Prototypes for Extended Metaverse Agent Systems}


The architecture in the previous section introduced the components needed to extend the metaverse and the kinds of connectivity and transitions necessary for users. To examine the possibility of extended metaverse agents for solving the disconnected problem between metaverse and physical space, this work instantiates the unique architecture into a prototyping framework. Two variations of this framework and their resulting prototypes are presented as proof-of-concepts.

The first prototype, MetaPlant (see Figure \ref{Prototypes}(a)), is to understand how extended metaverse agents can be designed to help users visualize their context while in an immersive metaverse, using computer vision modelling to show the temporal context and application context (i.e., what applications are they using). This exploration helps to show how the metaverse can be used in an engaging design that potentially impacts the user's experience. The second prototype, Meta-RV-Traveller(see Figure \ref{Prototypes}(b))  considers how an extended metaverse agent can be used when transitioning across the mixed reality spectrum (i.e., allowing the user to move from full Virtual Reality to Mixed Reality while understanding the general shared context). This uses computer vision to detect more objects of the surroundings and to provide visualization  allowing the user to remain aware of the physical context without actually taking them out of the experience. In both prototypes, the extended metaverse agent(s) is envisioned as taking on the role of an ``assistant'' that helps improve the connections between metaverse context and user physical context through mixed reality dynamics and visualizations. These are both described in Figure 1 \cite{guan2022extendingThesis}, including their overall framework and design outcomes.

\section{Summary}
This work provides an early investigation into the development of frameworks for a  stronger coupled metaverse with a focus on bridging virtual reality, mixed reality, and physical reality as a step toward extending the metaverse potential. It directly highlights the metaverse disconnect challenge, and offers a design-based architectural framework for exploring scenarios for these strongly connected metaverse environments. Two framework instantiations are presented, for rich connectivity and context visualizations, as well as for streamlining these spaces as users transition between them. Together, they provide an early proof-of-concept of what may be possible in the metaverse of the near future, with hybrid virtual-physical objects in an extended IoT-enabled metaverse smart environment.

\acknowledgments{
This work was supported in part by the Tri-council of Canada under the Canada Research Chairs program.}

\bibliographystyle{abbrv-doi}

\bibliography{template}
\end{document}